# Searching for Superconductivity in High Entropy Oxide Ruddlesden-Popper Cuprate Films

To be published in JVST-A special issue honoring Scott Chambers

Alessandro R. Mazza[1], Xingyao Gao[1], Daniel J. Rossi[2], Brianna L. Musico,[3] Tyler W. Valentine[2], Zachary Kennedy[2], Jie Zhang[1], Jason Lapano[1], Veerle Keppens[3], Robert G. Moore[1], Matthew Brahlek[1], Christina M. Rost[2], and Thomas Z. Ward[1,a)]

[1]Materials Science and Technology Division, Oak Ridge National Laboratory, Oak Ridge, TN, 37831, USA

[2]Department of Physics & Astronomy, James Madison University, Harrisonburg, VA 22807, USA

[3]Department of Materials Science and Engineering, University of Tennessee, Knoxville, TN 37996, USA

a) Electronic mail: wardtz@ornl.gov

**ABSTRACT**: In this work, the high entropy oxide $A_2CuO_4$ Ruddlesden-Popper $(La_{0.2}Pr_{0.2}Nd_{0.2}Sm_{0.2}Eu_{0.2})_2CuO_4$ is explored by charge doping with $Ce^{+4}$ and $Sr^{+2}$ at concentrations known to induce superconductivity in the simple parent compounds, $Nd_2CuO_4$ and $La_2CuO_4$. Electron doped $(La_{0.185}Pr_{0.185}Nd_{0.185}Sm_{0.185}Eu_{0.185}Ce_{0.075})_2CuO_4$ and hole doped $(La_{0.18}Pr_{0.18}Nd_{0.18}Sm_{0.18}Eu_{0.18}Sr_{0.1})_2CuO_4$ are synthesized and shown to be single crystal, epitaxially strained, and highly uniform. Transport measurements demonstrate that all as-grown films are insulating regardless of doping. Annealing studies show that resistivity can be tuned by modifying oxygen stoichiometry and inducing metallicity but without superconductivity. These results in turn are connected to extended x-ray absorption fine structure (EXAFS) results indicating that the lack of superconductivity in the high entropy cuprates likely originates from a large distortion within the Cu-O plane ($\sigma^2 > 0.015$ Å$^2$) due to A-site cation size variance, which drives localization of charge carriers. These findings describe new opportunities for controlling charge- and orbital-mediated functional responses in Ruddlesden-Popper crystal structures, driven by balancing of cation size and charge variances that may be exploited for functionally important behaviors such as superconductivity, antiferromagnetism, and metal-insulator transitions, while opening less understood phase spaces hosting doped Mott insulators, strange metals, quantum criticality, pseudogaps, and ordered charge density waves.

## I. INTRODUCTION

Since the discovery of high $T_c$ superconductivity in 1986[1], the underlying mechanism for Cooper pair formation in the cuprates has continued to baffle researchers. Both electron and hole doping from the antiferromagnetic Mott insulating ground state of the parent compounds leads to a rich phase diagram where pseudogaps, density waves and strange metals surround a superconducting dome[2]. Based on competing experimental evidence from a variety of techniques, two prominent schools of thought for the formation of the pairing glue have emerged. Inelastic neutron scattering[3,4] and Resonant Inelastic X-ray Scattering[5] (RIXS) have revealed an interplay between antiferromagnetic spin fluctuations and superconducting pairs with exchange energy scales sufficient for forming a superconducting condensate, suggestive of a spin mediated pairing mechanism[6–8]. On the other hand, "kinks" in the electronic dispersion revealed by Angle-Resolved Photoemission Spectroscopy[9] (ARPES) as well as the isotope effect in both ARPES[10] and Scanning Tunnelling Spectroscopy[11] (STS) measurements provide evidence for strong electron phonon coupling as the origins of the Cooper pairing glue[2,12]. Similar debates appear regarding the origins of the pseudogap phase in hole doped compounds and whether it is due to preformed Cooper pairs without phase coherence[13] or density wave phenomena competing with superconductivity[14,15]. Differences in the pseudogap phase are also observed with the electron doped compounds where AFM order is more robust[2]. Despite the array of conflicting information, the physical arrangement of the atomic layers is simplistic in that it can be viewed as alternating layers of ionic metal-oxygen charge reservoir layers interleaved with active doped $CuO_2$ layers where superconductivity is believed to reside. A key to unraveling the mysteries of high $T_c$ cuprate superconductors is finding new ways to disentangle the contributions of electrons, lattice, and spins in the electronic structure. In this study, we create a metal-oxide charge reservoir with high entropy, while leaving the doped $CuO_2$ layers intact. The random distribution of cations in the charge reservoir layer should provide a control parameter on the phononic and magnetic degrees of freedom in the system.

The extraordinary compositional and structural tunability of the high entropy oxides (HEOs) make them outstanding candidates to provide never before possible accessibility to the nearly degenerate spin, charge, orbital, and lattice energies that drive emergent correlated behaviors. HEOs are an emerging class of materials in which one or more cation sublattices host 5 or more different cations connected by an oxygen anion sublattice[16–18]. An important distinction of HEOs from more traditional metal-metal bonded high entropy alloys is the fact that the bonding network of the HEOs are built upon covalent and ionic bonds which permit structural motifs and electron interactions not accessible in metal-metal bonded high entropy alloys[19–22]. When comparing to less compositionally complex oxides, the presence of the many different cations in the HEOs act to facilitate uniform mixing on cation sublattices[16,23–25]. This allows for chemical doping design well beyond what is conventionally accessible, since materials of lower complexity are more strongly dominated by enthalpic energies during synthesis which leads to chemical segregation and formation of impurity phases[26]. Entropy stabilization has led to the synthesis of a broad array of functionally relevant crystal structures including spinel[27–29], rocksalt[30–32], Ruddlesden-Popper (RP)[33,34], and perovskite[26,35–37]. In the ideal case, the microstructure of HEOs can be viewed as the solid state of an ideal solution, which can bring about unique functionalities compared to conventional oxides. For example, HEOs have exhibited remarkable thermal[38–40], optical[19,41], and dielectric properties[42,43] with potential applications in thermal barrier coatings[38,39], Li-ion batteries[44,45], and catalysis[21]. Importantly, studies have shown that HEOs possess unexpected thermal transport behaviors, which point to the ability to manipulate phononic degree of freedom through cation selection[36,38]. Functionalities in strongly correlated HEO systems are still in their infancy, however recent works related to magnetism and the development of synthesis approaches that enable single crystal HEO films suggest that this area is likely to produce many exciting and unexpected behaviors, which are not accessible in low complexity systems.[20,27,28,30]

In this work, we investigate electron and hole doping of $(La_{0.2}Pr_{0.2}Nd_{0.2}Sm_{0.2}Eu_{0.2})_2CuO_4$ (5$A$CuO) – allowing both the exploration of the effects on cation size variance and electronic doping on the electronic state. Both n-type (via Sr) and p-type (via Ce) dopings are shown to be single-phase and epitaxial. Regardless of doping type, the as-grown films are insulating at all measured temperatures. Owing to the known sensitivity to oxygen mobility and tunability in Cu-O plane morphology[46,47], annealing studies are carried out on all samples. X-ray diffraction (XRD) measurements confirm characteristics of changing oxygen vacancy density that result in changes to resistivity. In the electron doped sample, a high temperature metallic state is achieved, but no evidence of superconductivity is observed. Extended x-ray absorption fine structure (EXAFS) measurements show buckling in the Cu-O plane which is suspected as the primary driver in preventing the onset of superconductivity. As oxygen vacancies form, buckling in the Cu-O plane decreases, which relieves local strain from different cation sizes and is the likely mechanism that induces the observed drop in resistivity after annealing in reducing environments. These results provide insight into the evolution of electronic phase in high entropy cuprates, where strain, oxygen vacancy density, and *A*-site cation size variance are all found to play a role in functionality.

## II. EXPERIMENTAL

### A. *Sample Growth*

All films are synthesized at 720 °C using pulsed laser deposition (PLD) with a 248 nm KrF excimer laser at a fluence of 0.7 J/cm$^2$ and frequency of 5 Hz with a steady oxygen pressure of 1 mTorr throughout growth and cooling to room temperature at a rate of 25 °C/min. Targets are prepared using the conventional solid-state method with a stoichiometric ratio of cation elements[34]. The mix of cations on the *A*-site sublattice each have an equivalent ternary component parent phase which have different preferred tetragonal (T) phases. As examples, $La_2CuO_4$ (LCO) is generally associated with the pure T-type structure with a 6-fold oxygen coordination around the Cu atoms; while $Nd_2CuO_4$ (NCO) tends toward the T'-type structure which is characterized by 4-fold Cu coordination in the plane. The type of charge doping required to generate superconductivity is often correlated to these structures with LCO requiring ~20% hole doping and NCO requiring ~15% electron doping. Previous studies on undoped 5*A*CuO show that the high entropy cuprate prefers the 2D-like T' structure[33]; however we synthesize both charge and hole doped samples at what would be considered optimal doping levels for both the LCO and NCO parents[48]. Thus, 5*A*CuO is doped with 15% $Ce^{4+}$ and 20% $Sr^{2+}$ in the forms $(La_{0.185}Pr_{0.185}Nd_{0.185}Sm_{0.185}Eu_{0.185}Ce_{0.075})_2CuO_4$ (abbreviated as Ce-5*A*CuO) and ($La_{0.18}Pr_{0.18}Nd_{0.18}Sm_{0.18}Eu_{0.18}Sr_{0.1})_2CuO_4$ (abbreviated as Sr-5$A$CuO). The samples are grown on $SrTiO_3$ (STO) (001) and $DyScO_3$ (DSO) (110) substrates.

### B. Structural Characterization

The thickness and microstructure of the as-deposited films are characterized via x-ray diffraction (XRD) using a PANalytical X'Pert PRO, X'Pert[3] MRD equipped with Cu K$\alpha_1$. *In-situ* high temperature XRD measurements are conducted in air using a heating stage. Small background peaks observed in these scans result from the Inconel heating stage, the polymer dome/heat shield, and the silver epoxy used in mounting the sample. Atomic Force Microscopy (AFM, Digital Instruments) measurement using tapping mode are conducted to characterize the film surface.

The Cu environment in thin film samples of 5$A$CuO and Ce-5$A$CuO grown on STO substrates are measured using EXAFS. K-edge (8999 eV) spectra are collected at beamlines 12-BM-B and 10-BM-B[49] at the Advanced Photon Source, Argonne National Laboratory (Lemont, IL), using Canberra 13-element and Vortex fluorescence detectors, respectively. Between 6-11 individual scans of each sample are collected within the energy range -200 to 900 eV above the absorption edge, then processed and analyzed using the Demeter package[50]. Scans are corrected for self-absorption using the Fluo[51] algorithm, normalized, and merged resulting in one scan for each sample. Each scan is calibrated in energy using a Cu reference foil measured in sequence with the films.

A Fourier transform (FT) is performed on background subtracted $\chi(k)$ spectra using a Hanning window range of 2.90 ≤ 12.45 Å$^{-1}$. Given the resolution of the resulting partial pair distribution function is related to the k-range by $\pi/(2\Delta k)$, the real space resolution limit is ~0.17 Å. Quantitative analysis of the local structure around Cu is accomplished by fitting up to the first six coordination shells to the phase-uncorrected FT, from 1-4.3 Å. Our fitting model consists of T' phase unit cell with in-plane lattice parameter of 3.905 Å, matching the STO substrate. The out-of-plane lattice constant is taken as the average between that of Sr-5$A$CuO (12.22 Å) and Ce-5$A$CuO (12.08 Å)), to obtain c = 12.145 Å. The lattice parameter of 5$A$CuO (12.21 Å) falls within this range as well. Average occupancy of the $A$-site is divided evenly among five lanthanides: La, Pr, Sm, Nd, and Eu to match sample composition. Ce and Sr are introduced using $Ce_2CuO_4$ and $Sr_2CuO_4$, respectively, with modified lattice constants to generate additional scattering paths in FEFF[52] and average occupancies are modified to account for the compositional change on the $A$-site. Up to six parameters per coordination shell for both data sets from the anharmonic (cumulant expansion) EXAFS equation[53]: the average coordination ($N_0$), the inner potential energy shift ($E_0$), the half scattering-path length (R), the EXAFS Debye-Waller factor ($\sigma^2$), and the third cumulant ($\sigma^3$). $S_0^2$, the amplitude reduction factor, was held constant for all data sets. $E_0$ is constrained based on beamline – meaning both undoped and Ce-doped samples are the same given they were measured on 12-BM, and Sr-doped was slightly different as it was measured on 10-BM. Other variables were free to float resulting in a cumulative total of 37 variables used out of 59 available points. R, $\sigma^2$, and $\sigma^3$ represent average values per coordination shell, encompassing homogenous cation distributions. $N_0$ represents the average coordination number of oxygen nearest neighbors.

### C. *Electronic Characterization*

Van der Pauw configuration is utilized for transport measurements in a Physical Property Measurement System (PPMS, Quantum Design). During transport measurements, the samples are zero-field cooled from 300 K to 10 K with a cooling rate of 5 K/min.

## III. RESULTS AND DISCUSSION

The structure of the as-deposited films is first examined by XRD. **Fig. 1(a)** shows the specular diffraction patterns of the undoped, Ce-5$A$CuO and Sr-5$A$CuO films grown on STO (001) substrates. XRD demonstrates that the films are of single crystal RP structure with preferred (00$l$) growth direction and no secondary phases[33]. The Ce and Sr doped samples exhibit differences in the c-axis lattice parameter, which likely originates from the different cation radius of the dopants. Specifically, $c = 12.08$ Å for Ce-5$A$CuO, $c = 12.22$ Å for Sr-5$A$CuO, and $c = 12.21$ Å for the native 5$A$CuO with the contraction/expansion of the doped films being consistent with the different ionic radii of the dopants, though it should be noted that the relative oxygen vacancy densities of the films also have some impact on lattice parameters. A reciprocal space mapping (RSM) scan of the Ce-5$A$CuO film (1 0 11) taken near the STO (1 0 3) peak is shown in Fig. 1(b), where the film is shown to be epitaxially strained to the substrate. The electronic character of each of the films are characterized using transport measurements. As seen in 1(c) all of the as-grown films are insulating in nature, with the Sr-doped film being immeasurable in its virgin state.

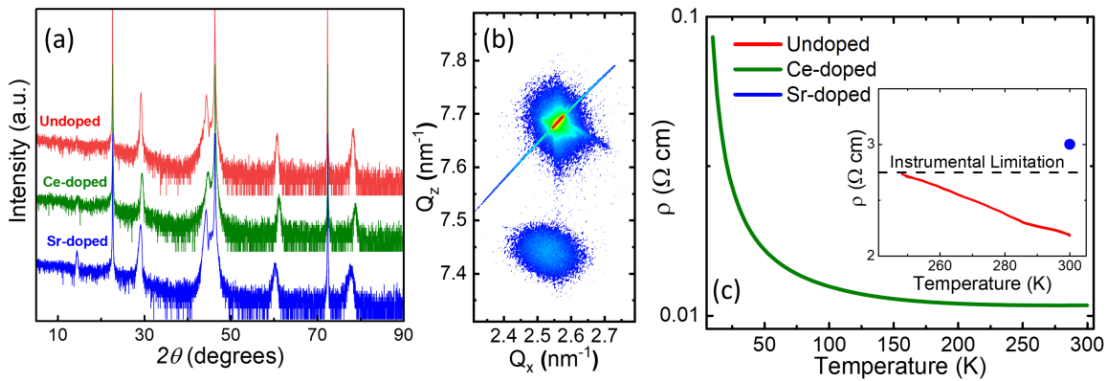

**Fig. 1** (a) θ-2θ XRD comparison of the undoped, Ce doped, and Sr doped 5$A$CuO thin films on STO (0 0 1) substrates. (b) RSM scan of the Ce-5$A$CuO (1 0 11) near the STO (1 0 3) peak. (c) Resistivity of as-grown films. The Sr-doped composition exceeds measurable resistivity in the as-grown state.

After characterizing the as-grown state of the films, annealing studies are conducted on all films to better understand how insertion and removal of oxygen influences structure and behavior. This is important, since modifying oxygen vacancy density is known to influence electronic character in the parent cuprate systems[47]. The resulting *in-situ* high temperature XRD of a Ce-5$A$CuO film is shown in **Fig. 2(a)**. Here the sample was heated from 300 °C to 850 °C in air where remarkably little change is observed. The film shows high thermal stability even at temperatures higher than its deposition temperature of 720 °C, where the (004) and (006) film diffraction peaks show no significant change in intensity and no extraneous peaks appear. A small expansion is observed with increasing temperature, which is consistent with thermal expansion of the film.

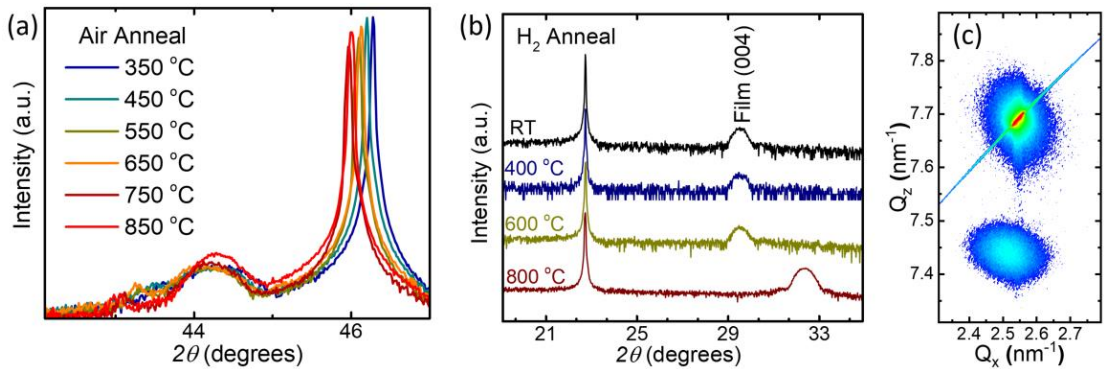

**Fig. 2 (a)** Heating XRD scan of the Ce-5$A$CuO thin film at the RP (006), from 350 °C to 850 °C in air. The small peak ~43° is due to the heating stage, see experimental section for details. (b) XRD comparison of the Ce-5$A$CuO thin film after annealed in H$_2$ at different temperatures. (c) RSM scan of the Ce-5$A$CuO thin film near the STO (103) peak after H$_2$ annealing at 400 °C.

Establishing that little to no change occurs upon annealing in air, additional experiments annealing in ozone (10% O$_3$/90% O$_2$) and H$_2$ (forming gas, 5% H$_2$/95% N$_2$) were done to further promote the insertion/removal of oxygen respectively. Fig. 2(b) shows the resulting XRD of the film (004) for a Ce-5$A$CuO film after annealing at different temperatures in H$_2$. A small expansion in the out-of-plane lattice parameter is observed with increased annealing temperature up to 600 °C; a change which is consistent with loss of oxygen. Despite the increased temperature, the film's crystal structure is well maintained with clear RP phase and no impurity peak formation or shifts to in-plane registry (Fig. 2(c)). At higher temperatures of 800 °C, a dramatic peak shift is observed, which indicates that the film undergoes a phase transition. This is accompanied by a color change of the sample, where the film is converted from dark to transparent and is commensurate with the formation of mobile oxygen vacancies, as discussed in more detail below. Additional XRD from $O_3$ annealed Ce-5$A$CuO and the Sr-5$A$CuO films (after annealing in $H_2$ and $O_3$) can be found in the supplementary materials (Fig. S1)[54].

Transport properties of doped 5$A$CuO films are investigated as a function of annealing. **Fig. 3(a)** shows the resistivity vs. temperature curves of a Ce-5$A$CuO film after annealing at various temperatures in $H_2$. After 200 °C annealing in $H_2$, the film's resistivity is relatively unchanged, as the thermal threshold for oxygen migration has not been met. As annealing temperature increases, the resistivity of the film decreases over the entire temperature range. The film resistivity at room temperature is decreased by about 20% after annealing at 400 °C and further decreased by 50% after the 600 °C anneal. Importantly, a high temperature metallic phase appears (Fig. 3(b)) after forming gas annealing at 400 °C, which demonstrates that not only resistivity, but also electronic phase transitions can be controlled through oxygen stoichiometry. After annealing the sample at 800 °C, the film becomes immeasurably insulating, which coincides with the structural changes observed in XRD. This, considering the notable change in XRD, may be indicative of a critical oxygen vacancy density at which a vacancy ordered phase might emerge, though further microscopy and x-ray diffraction should be done to confirm.

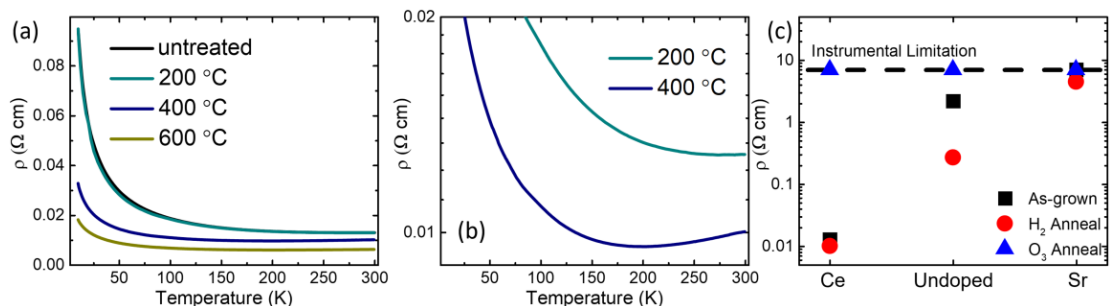

**Fig. 3** Annealing effects on Ce-doped films. (a) Temperature dependent resistivity comparison of the Ce-5$A$CuO thin film on STO (001) substrate after annealed in $H_2$ at different temperatures. The result of 800 °C was not plotted due to the huge resistance that exceeded the measurement limit of the equipment. (b) Resistivity for the 200 °C and 400 °C showing the emergence of a MIT ~200K. (c) Resistivity at 25 °C (300K) for samples in the as-grown, $O_3$ annealed, and $H_2$ forming gas (at 400 °C) annealed states.

In addition to the $H_2$ annealing process aimed at reducing oxygen content, the Ce-5$A$CuO films are also annealed in $O_3$ with the intent of increasing oxygen content from the as-grown state. Fig. S1(a) shows the XRD pattern of a Ce doped film after annealing in $O_3$ at 200 °C. The film shows no obvious structure change and no new phase formations. However, the transport properties of the film are greatly changed via an intense increase (beyond the limitation of available instrumentation) in the resistivity, resulting from the decrease of intrinsic oxygen vacancies. The same annealing processes were conducted on undoped and Sr-doped films. In combination with $H_2$ annealing experiments, these results show that the formation of oxygen vacancies lead directly to improved transport properties. These results are summarized in Fig. 3(c), where all films are also shown to be immeasurable after annealing in ozone. To understand the mechanism of the observed functional changes, it is important to first rule out possible extrinsic effects that could result from the STO substrate becoming more conducting with oxygen vacancies[55]. This effect is confirmed to be intrinsic to the film by growing on DSO substrates, as detailed in the supplementary materials (Fig. S1, S2)[54].

One possible intrinsic explanation for the change in transport is that the shift in oxygen content simply acts to donate electrons to the copper charge carrier concentration. However, this is unlikely to be the only factor. Changes to Cu-O plane bond distortions driven by cation size variation on the perovskite layer are well known to influence electron conduction in cuprates[56,57]. Furthermore, the cation disorder may enable the local trapping of charge which is then mediated by the rearrangement of oxygen vacancies which help redistribute charge resulting in a drop in resistivity. This concept regarding the polaronic tendencies of these materials can be further investigated by, for example, ARPES[2]. In the present study, we investigate Cu-O plane distortions which would influence inherent stable oxygen coordination environments after initial synthesis. This would partially contribute to the large difference in resistivities of the as-grown samples. The changes to resistivity observed in the undoped and Sr doped samples after annealing would then be influenced by the reduction in buckling of the Cu-O plane due to changes to oxygen coordination. To investigate this further, it is important to identify both the Cu coordination and local displacement in the undoped film and doped films.

X-ray absorption near edge structure (XANES) was used to study the electronic and valence state of Cu. Normalized K-edge XANES of the Cu K-edge for 5*A*CuO, Ce-5*A*CuO, and Sr-5*A*CuO are plotted in **Fig. 4(a)** and compared to a Cu metal foil for reference. The edge was consistently chosen across all spectra to be the first peak in $\partial\mu/\partial E$. The primary feature changing between the three films is a slight shift to lower or higher energy with the introduction of Ce or Sr, respectively. In the case of Ce-5*A*CuO this suggests a partial change in oxidation state from $Cu^{2+}$ to $Cu^{1+}$, however the actual ratio of these states cannot be determined without further measurement. In the Sr-5*A*CuO this suggests a dominant $Cu^{2+}$ state. Further analysis, comparing measured spectra to a reference spectrum from previous studies[33,58] (Fig. 4(b)), suggests all films are square planar coordinated in the T' phase. Having established the T' phase and expected modification of the Cu valence, we turn to the structure of the cuprates to understand the implication cation disorder may have on the conducting Cu-O plane.

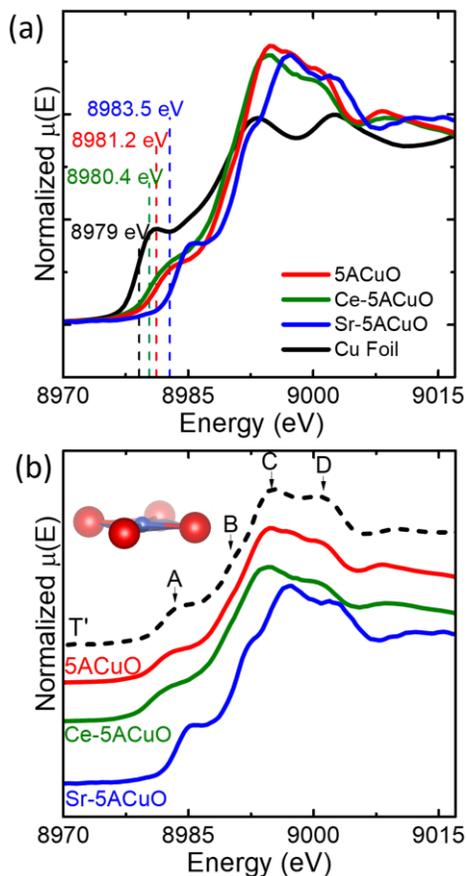

**Fig. 4** Normalized Cu-K edge XANES spectra comparing 5*A*CuO and Ce-5*A*CuO to (a) the Cu calibration foil and (b) reference spectra depicting key features of the T' cuprate phase from a previous report[33,57].

A key in the conductivity of cuprates is the fine structure, where inhomogeneity in the Cu-O plane can lower the critical temperature of a superconducting or metallic phase[59,60]. This has been similarly observed as a function of *A*-site cation size variance, where increasing variance leads to decreased $T_C$[57,61]. With regards to the precluded metallic phase in 5*A*CuO, EXAFS for both undoped and doped (Ce and Sr) compositions were done to determine how *A*-site cation disorder affects the conducting Cu-O plane. **Fig. 5(a-c)** show the best-fit magnitude ($|I\chi(R)|$) and imaginary part ($\chi_{im}(R)$) of the k-weighted FT of the EXAFS spectra for 5*A*CuO, Ce-5*A*CuO, and Sr-5*A*CuO respectively. The final fit results and additional details for each composition are provided in the supplementary materials (Table S1)[54]. The half scattering-path length, R, varies appropriately with the unit cell dimensions of each sample composition, and further strengthens confidence in our fit. Fits for shells with scattering paths that have a component corresponding to the out-of-plane lattice parameter should differ according to the actual lattice parameters of the sample, as opposed to the average model value. It can be seen that this is the case as the second, third, and fifth shells all have consistently smaller values for Ce-5*A*CuO, which has a smaller out-of-plane lattice parameter compared to 5*A*CuO and Sr-5*A*CuO.

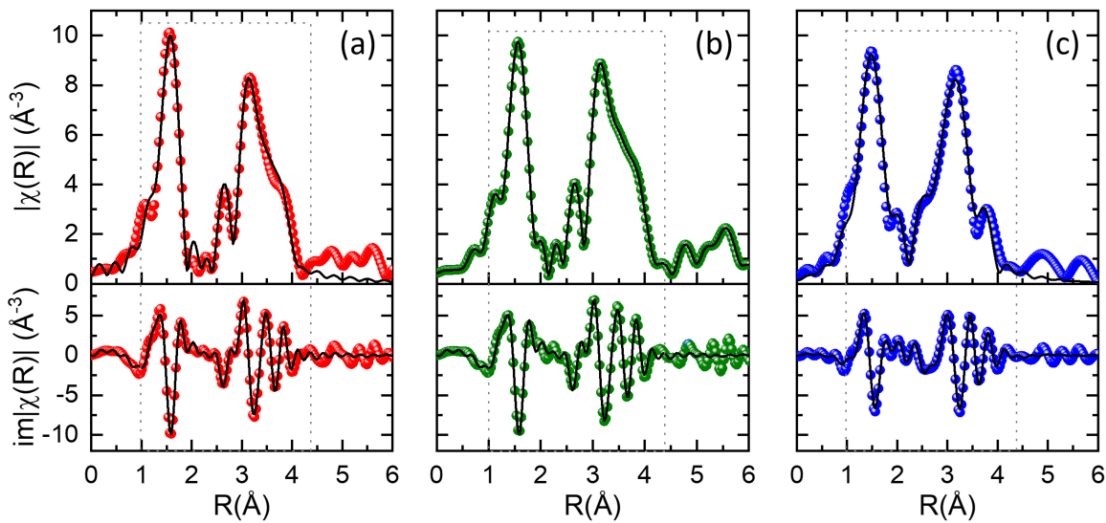

**Fig. 5** Best fit magnitude and imaginary parts, denoted by the dashed boundaries, of the EXAFS Fourier transform for (a) 5*A*CuO, (b) Ce-5*A*CuO, and (c) Sr-5*A*CuO.

EXAFS provides key insight into the prevention of a superconducting or even metallic phase in 5*A*CuO and its doped counterparts. First, of note is the substantially lower coordination number of oxygen in the Sr-5*A*CuO sample relative to each of the other compositions. This likely corresponds to oxygen vacancies, indicating the Sr-doped samples (as-grown) are inherently more disordered perhaps contributing to their more insulating behavior relative to the undoped and Ce-doped compositions. Exact mobility and oxygen vacancy densities as a function of annealing and doping are the subject of ongoing research and will provide further insight to this perceived effect. Next, the presence of cation disorder on the *A*-site appears to lead to a large RMS distortion (EXAFS

Debye-Waller, $\sigma^2$) of the Cu-O plane, regardless of dopant, which has been shown to suppress conductivity in other cuprates[59,60]. This agrees with studies showing that a large variance in cation size negatively impacts the critical temperature of conducting perovskites and cuprates[56,62] The effect of the formation of oxygen vacancies, rather than a simple modification to the electronic structure, is shown to have a minimal impact on conductivity with traditional hole/electron doping in the present work. Therefore, it is believed the primary impact of annealing is on the structural homogeneity of the films. As in the case of strain engineering, oxygen vacancy formation is known to directly affect the Cu-O lattice plane homogeneity. When oxygen vacancies form in doped 5*A*CuO, we suggest the large RMS roughness is decreased in the Cu-O plane, by vacancies forming selectively near highly distorted regions, which in turn decreases resistivity in this plane.

## IV. SUMMARY AND CONCLUSIONS

The effects of hole (Sr) and electron (Ce) doping on the high entropy cuprate 5*A*CuO were investigated via x-ray scattering and transport experiments. XRD demonstrates the PLD grown films are single phase and epitaxially strained to the substrate. From in-situ XRD it is found that the films exhibit excellent thermal stability, showing no structural change when annealing to 850 °C in air and 600 °C in $H_2$. Each of the films, regardless of doping, are found to be insulating at all measured temperatures. However, the resistivity of the films could be effectively tuned by annealing in a reducing (forming gas) or oxidizing (ozone) environment. Increased oxygen vacancy density in the films was shown to effectively decrease the resistivity of films for all *A*-site doping, with a MIT emerging for the Ce-5*A*CuO film. While doping shows limited change to transport properties, XANES confirms Cu valence changes with doping in the expected manner.

From EXAFS, we gain insight into the mechanism preventing a metallic or superconducting phase in the as-grown state as a large distortion of the Cu-O plane is observed. This distortion has been connected to lowering critical temperatures in cuprates and observed in systems where *A*-site cation size variance is attributed to decreases in $T_c$. In connection to transport results, the EXAFS suggests the primary role of oxygen vacancy formation in increasing conductivity may be in lowering the distortion or buckling of the Cu-O plane. These results shed light into a possible mechanism towards finding superconductivity in entropy stabilized cuprates, where a delicate balance of cation size variance, oxygen vacancy density, and strain must be considered as they relate to the uniformity of the conducting Cu-O plane. While superconductivity was not observed in any films presented in this work, it appears that superconductivity is not out of reach in these materials with tuning of size variances to reduce distortion in the Cu-O plane.

**ACKNOWLEDGMENTS**

Experiment design, sample synthesis, structural characterization, and computational modelling were supported by the US Department of Energy (DOE), Office of Basic Energy Sciences (BES), Materials Sciences and Engineering Division. This research used resources of the Advanced Photon Source, a U.S. Department of Energy (DOE) Office of Science User Facility operated for the DOE Office of Science by Argonne National Laboratory under Contract No. DE-AC02-06CH11357. CMR, DJR, TWV, and ZK gratefully acknowledge support from NSF DMR-2011839. CMR also expresses gratitude to Benjamin Reinhart, APS sector 12, and Mark Warren and Joshua Wright, APS sector 10, for mail-in assistance on measurements during the COVID-19 pandemic. RGM acknowledges support by the U.S. Department of Energy, Office of Science, National Quantum Information Sciences Research Centers, Quantum Science Center. B.L.M. acknowledges the support from the Center for Materials Processing, a Tennessee Higher Education Commission (THEC) supported Accomplished Center of Excellence.

**DATA AVAILABILITY**

Data contained in this work made available upon reasonable request by contacting the corresponding author.